\documentclass[twocolumn,nofootinbib,jcp,preprintnumbers,amsmath,amssymb]{revtex4-1}
\usepackage{natbib,hyperref}
\usepackage{amsmath}
\usepackage{stmaryrd}
\usepackage{braket}
\usepackage{graphicx}

\usepackage{dcolumn}
\usepackage{tabularx}
\usepackage{bm}
\usepackage{epsfig,color,xspace,multirow,xr,bbold}
\usepackage[all]{xy}
\usepackage{setspace}
\usepackage{url}
\usepackage{threeparttable, tablefootnote}
\newcommand{\RRef}[1]{\,Ref.\nocite{#1}\citenum{#1}} 
\newcommand{\bigqm}{bigQM7$\rm{\omega}$\,} 
\usepackage{natbib,hyperref}
\usepackage{float}
\usepackage{placeins}
\usepackage[utf8]{inputenc}

\usepackage[none]{hyphenat}
\usepackage{calligra}

\begin{document}
% \preprint{AIP/123-QED}

\title[]{Resolution-vs.-Accuracy Dilemma in Machine Learning Modeling of Electronic Excitation Spectra}

\author{Prakriti Kayastha}
\thanks{These authors contributed equally to this work.}
\author{Sabyasachi Chakraborty}
\thanks{These authors contributed equally to this work.}
\author{Raghunathan Ramakrishnan}
\email{ramakrishnan@tifrh.res.in}
\affiliation{Tata Institute of Fundamental Research Hyderabad, Hyderabad 500046, India.}

\date{\today}

\begin{abstract}
In this study, we explore the potential of machine learning for modeling molecular electronic spectral intensities as a continuous function in a given wavelength range. Since presently available
chemical space datasets provide excitation energies and corresponding 
oscillator strengths for only a few valence transitions, here, we present a new dataset---\bigqm---with 12,880 molecules containing up to 7 CONF atoms and report ground state and excited state properties.
A publicly accessible web-based data-mining platform is presented to facilitate on-the-fly screening of 
several molecular properties including harmonic vibrational and electronic spectra.
We present all singlet electronic transitions from the ground state calculated using
the time-dependent density functional theory framework with the $\omega$B97XD exchange-correlation
functional and a diffuse-function augmented basis set.
The resulting spectra predominantly span the X-ray to deep-UV region (10--120 nm).  
To compare the target spectra with predictions based on small basis sets, we bin spectral intensities and show 
good agreement is obtained only at the expense of the resolution.
Compared to this, machine learning models with latest structural representations trained 
directly using $<10 \%$ of the target data recover the spectra of the remaining molecules 
with better accuracies at a desirable $<1$ nm wavelength resolution. 
\end{abstract}

\keywords{
Crystallography,
Materials,
Phonons,
Supercell,
Wyckoff positions
}

\maketitle

\section{Introduction}\label{sec_intro}
The future of chemistry research hinges on the progress in data-driven autonomous discoveries\cite{steiner2019organic,christensen2021data,stach2021autonomous}. 
The performance of intelligent infrastructures necessary for such endeavors, such as chemputers\cite{gromski2020universal}, can be tremendously enhanced when augmenting experimental data used for their training with accurate {\it ab initio} results\cite{li2021combining,bai2019accelerated}. 
For designing opto-electronically important molecules such as dye-sensitized solar cells\cite{mathew2014dye,abreha2019virtual}, sunscreens\cite{sampedro2011computational,losantos2017rational}, or organic photovoltaics\cite{hachmann2011harvard,pyzer2015learning}, the corresponding target properties are excitation energies and the associated spectral intensities. %Referee 3, major point 2
Additionally, successful molecular design also requires information about thermodynamic/dynamic/kinetic stabilities, molecular lifetimes, solubility, and other experimental factors pertaining to molecular characterization.
Accelerated discoveries based on  molecular design workflows require a seamless supply of accurate theoretical results. To this end, machine learning (ML) models trained on results from {\it ab initio} predictions have emerged as their rapid and accurate surrogates\cite{rupp2012fast,ramakrishnan2017machine,von2018quantum}.
%Referee 3, major point 1

\begin{figure*}[!htp]
    \centering
    \includegraphics[width=15cm]{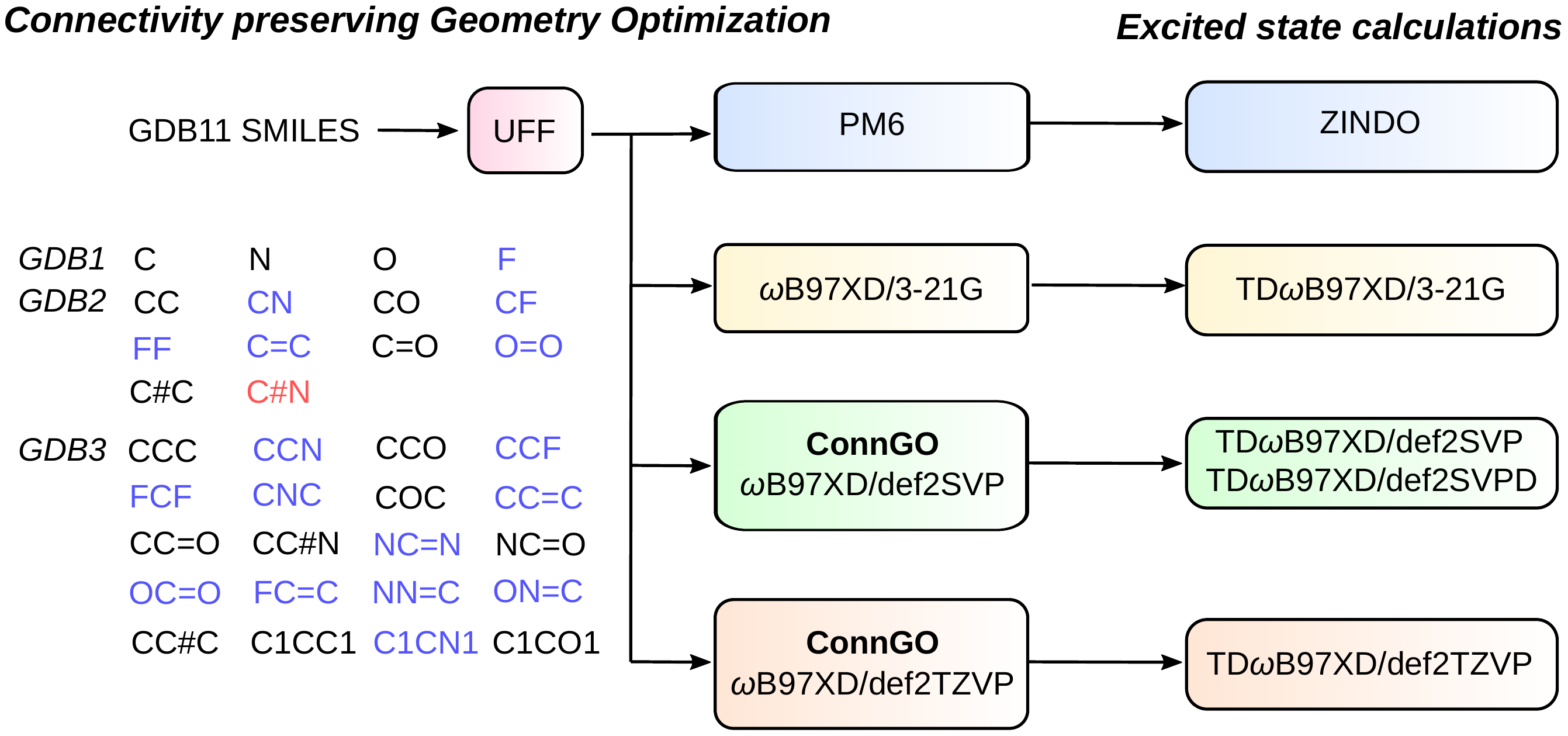}%\linewidth
    \caption{\bigqm chemical space design: Molecular composition and data generation workflows. From the GDB11 dataset, SMILES descriptors for all molecules with up to 7 CONF atoms were collected. For the GDB1-GDB3 subsets, CHONF molecules present in GDB11 but absent in GDB17 are shown in blue. 
    HCN that is present in GDB17, but absent in GDB11 is shown in red. Initial geometries were obtained with UFF that are subsequently optimized at the PM7 and $\omega$B97XD/3-21G levels. $\omega$B97XD geometry optimizations with large basis sets (def2SVP and def2TZVP) were done using the ConnGO workflow. TDDFT single point calculations were done using the DFT equilibrium geometries, while ZINDO calculations were done at PM6 geometries.}
    \label{fig:my_label}
\end{figure*}
% \textcolor{blue}{In this study, QML\textcolor{green}{ML} stands for data-driven modeling where conventional statistical methods are trained on quantum chemistry big data of molecules/materials.} 
ML models have been shown to accurately forecast a multitude of global\cite{rupp2012fast,ramakrishnan2015many,ramakrishnan2015big} and quasi-atomic molecular properties\cite{rupp2015machine,gerrard2020impression,gupta2021revving}. 
%mark 2
For atomization or bonding energies, their prediction uncertainties are comparable to that of hybrid density functional theory (DFT) approximations\cite{hansen2015machine,ramakrishnan2017machine,faber2018alchemical,qiao2020orbnet,unke2019physnet,schutt2017schnet,faber2017prediction}. 
They have also successfully modeled non-adiabatic molecular dynamics\cite{prezhdo2021modeling}, vibrational spectra\cite{lam2020combining,kananenka2019machine}, electronic coupling elements\cite{caylak2019evolutionary}, excitons\cite{vela2021learning}, electronic densities\cite{grisafi2018transferable}, excited states in diverse chemical spaces\cite{ramakrishnan2015electronic,tapavicza2021elucidating,caylak2021machine}, as well as excited-state potential energy surfaces (PES)\cite{westermayr2020deep,westermayr2020machine,tapavicza2021elucidating,dral2021molecular,rankine2021progress}.
A key difference in the performance of ML in the latter two application domains is that 
ambiguities due to atomic indices and size-extensivity that affect the quality of 
structural representations for chemical space 
explorations\cite{hansen2013assessment,von2015fourier} do not arise in PES modeling or dipole surface modeling\cite{huang2005ab,behler2015constructing,manzhos2020neural}
resulting in better learning rates. 

ML models of global molecular energies (atomization/formation energies, etc.) with a robust structural representation benefit from the 
well-known mapping between the ground state electronic energy and the corresponding
minimum energy geometry established by the Hohenberg--Kohn theorem\cite{hohenberg1964inhomogeneous}.
The Runge--Gross theorem provides a similar mapping between the time-dependent potential and
the time-evolved total electron density\cite{runge1984density}. 
However, the target quantities in ML modeling of excited states are state-specific stemming from local molecular regions. %SC: Referee 5 has a point on exploring atomic SLATM's performance. However, the target quantities in ML modeling of excited states are state-specific, stemming from local molecular regions. 
For quasi-atomic properties such as $^{13}$C NMR shielding constants\cite{rupp2015machine,gerrard2020impression,gupta2021revving,rankine2021progress} or K-edge X-ray absorption spectroscopy\cite{rupp2015machine,rankine2021progress}, a representation encoding the local environment of the query atom results in better learning rates. 
Similarly, quasi-particle density-of-states---interpreted as intensities in a photo-emission spectrum---have also been successfully modeled\cite{mahmoud2020learning,westermayr2021physically}. 
However, for valence electronic excitations that are also local, the corresponding molecular substructure varies non-trivially across the chemical space. 
Hence, intensities based on oscillator strengths derived from many-electron excited state wave functions obeying dipole selection rules exhibit slow learning rates\cite{tapavicza2021elucidating}.
Determining the characteristic chromophore responsible for the electronic excitations is non-trivial for chemical space datasets such as QM9\cite{ramakrishnan2014quantum} that exhibit large structural diversity. 
This complexity, in turn, hinders the development of local descriptors that can map to the composition or structure of the chromophore and its environment.
Hence, we are limited to using global structural representations for ML modeling of electronic excited state properties. 
This limitation becomes evident from the modest performances of ML models of excitation 
energies\cite{ramakrishnan2015electronic,tapavicza2021elucidating}, and their zero-order approximations, the frontier molecular orbital (MO) energies\cite{faber2018alchemical,liu2021transferable,mazouin2021selected,caylak2021machine}. 
%Such representations while successfully mapping to most structure-dependent properties, do not always adequately represent the excitation paradigm. 
%This is because in the latter context, ground-state structure, excitation, and excited-state structure play important roles while the global structural representation only encodes the ground-state structure.

% \textcolor{blue}{For the chemical space datasets
% with large structural diversity, the characteristic chromophore responsible for the electronic excitations across molecules is not easy to determine. Hence, a local descriptor capturing the composition or structure of the chromophore and its environment is not feasible as long as the dataset is structurally diverse. % Referee 3, minor point 1
% To this end, ML modeling of electronic excited state properties is performed using global structural representations. 
% Hence, such models suffer from an information overload ({\it i.e.,} poor signal-to-noise ratio) in the feature space amounting to weak structure-property mapping.} This effect manifests in unsatisfactory performances of ML models for excitation energies\cite{ramakrishnan2015electronic,tapavicza2021elucidating}, and their zero-order approximations, the frontier molecular orbital (MO) energies\cite{faber2018alchemical,liu2021transferable,mazouin2021selected}.
\begin{table*}[!htp]
\centering
\begin{threeparttable}
        \caption{Comparison of volume, variety and veracity of selected small molecules
        chemical space datasets. Size, composition and methods (only DFT or post-DFT) 
        used for data generation are listed. 
        }
    \begin{tabular}{lllll}
    \hline 
     Details    & QM7 & QM7b & QM9${\tnote{a}}$ & \bigqm\\
         \hline
    Origin & GDB13 & GDB13 & GDB17 & GDB11 \\
    Elements & CHONS & CHONSCl & CHONF & CHONF \\
    Size & 7165 & 7211 & 133885 & 12880 \\
    Geometry optimization  & PBE0/tight tier-2 & PBE/tight tier-2 & B3LYP/6-31G(2{\it df},{\it p})& $\omega$B97XD/3-21G \\ 
     & &&& $\omega$B97XD/def2SVP \\
     & &&& $\omega$B97XD/def2TZVP \\
    Frequencies &  &  & B3LYP/6-31G(2{\it df},{\it p}) & $\omega$B97XD/3-21G \\ 
     & &&& $\omega$B97XD/def2SVP \\
    & &&& $\omega$B97XD/def2TZVP \\
    Excited states &  & $E_1$ & $E_1$, $E_2$, $f_1$, $f_2$ & all states \\
       &   & GW/tight tier-2 & RICC2/def2TZVP &  TD$\omega$B97XD/3-21G \\
       &   &  & TDPBE0/def2SVP  &   TD$\omega$B97XD/def2SVP\\
       &   &  & TDPBE0/def2TZVP  &  TD$\omega$B97XD/def2TZVP \\
       &   &  & TDCAMB3LYP/def2TZVP & TD$\omega$B97XD/def2SVPD  \\
    \hline 
    \end{tabular}
\begin{tablenotes}\footnotesize
\item [a] Contains 3993/22786 molecules with up to 7/ 8 CONF atoms. Excited state data are available for the 22786 subset\RRef{ramakrishnan2015electronic}.
\end{tablenotes}
    \label{datasettable}
    \end{threeparttable}
\end{table*}

In this study, we: 
(i) Present a high-quality chemical space dataset, \bigqm, containing ground-state properties and electronic spectra of 12,880 molecules containing up to 7 CONF atoms
modeled at the $\omega$B97XD level with different basis sets. 
(ii) Demonstrate the resolution-vs-accuracy dilemma in modeling spectroscopic intensities. 
(iii) Present ML models trained on the \bigqm dataset for an accurate reconstruction of the electronic spectra of allowed transitions in a given wavelength domain.

%SC: Present ML models trained on $\omega$B97XD/def2SVPD targets and demonstrate how these machines may be used to accurately reconstruct the full electronic spectra for entries in the \bigqm dataset.

\section{Chemical Space Design}
\subsection{The \bigqm dataset\label{sec_dataset}}
Pioneering efforts in small molecular chemical space design have culminated in the graph-based generated dataset, GDB11\cite{fink2005virtual,fink2007virtual}, containing 0.9 billion molecules with up to 11 CONF atoms. 
GDB11 provides simplified-molecular-input-line-entry-system (SMILES) string-based descriptors encoding molecular graphs. 
Larger datasets, GDB13\cite{blum2009970} and GDB17\cite{ruddigkeit2012enumeration} have since been created containing 13 and 17 heavy atoms, respectively. 
Synthetic feasibility and drug-likeness criteria eliminated several molecules in GDB13 and GDB17. 
Starting with the SMILES descriptors of GDB13, QM7\cite{rupp2012fast} and
QM7b\cite{montavon2013machine} quantum chemistry datasets emerged, provisioning computed equilibrium geometries and several molecular properties. 
Recently, QM7 has been extended by including non-equilibrium geometries for each molecule resulting in QM7-X\cite{hoja2021qm7}. Similarly, the QM9 dataset\cite{ramakrishnan2014quantum} used SMILES from the GDB17 library reporting structures and properties of 134k molecules with up to 9 CONF atoms.

In the present work, we explore molecules with up to
7 CONF atoms. We begin with the GDB11 set of SMILES because 
several important molecules such as ethylene and 
acetic acid present in GDB11 were filtered out in GDB13 and GDB17. 
Our new dataset contains 12883 molecules---almost twice as large as the QM7 sets. 
The breakdown for subsets with 1/2/3/4/5/6/7 heavy atoms is 4/9/20/80/352/1850/10568. 
The previous datasets QM7, QM7b, and QM9 have been generated using yesteryear's quantum chemistry workhorses: PBE\cite{perdew1996generalized}, PBE0\cite{adamo1999toward}, and B3LYP\cite{becke1993new}. 
Here, we use the range-separated hybrid DFT method, $\omega$B97XD\cite{chai2008long} that
is gaining widespread popularity for its excellent accuracy. 
Hence, we name this dataset as \bigqm, with the last character emphasizing the DFT approximation utilized. 
The high-throughput workflow used for generating \bigqm is shown in Fig.~\ref{fig:my_label} and
Table~\ref{datasettable} puts the new dataset in perspective by comparing with other popular datasets of similar constitution. While \bigqm is smaller than QM9, it 
provides a better coverage of molecules for the same number of CONF atoms. Further, \bigqm
also provides excited state data collected at various theoretical levels, hence, comprehensively
covering the property domain. A summary of properties of \bigqm, made available in the form of structured datasets\cite{kayastha2021Supplementary}, is provided in Table~\ref{structureddata}. 
%\textcolor{green}{w/o-} A consolidated summary of dataset size, composition, and theoretical levels used for data generation is presented in Table~\ref{datasettable}. 
As unstructured datasets, we provide
raw input/output files\cite{kayastha2021bigQM7w} to kindle future endeavors. For example, for ML modeling of forces, properties of non-equilibrium geometries may be extracted from these raw data.
\begin{table*}[!hbtp]
\begin{threeparttable}
    \centering
        \caption{Structured content of the \bigqm dataset\cite{kayastha2021Supplementary}. 
        %At various quantum chemistry levels, properties available are listed along with units.
        }
    \begin{tabular}{l}
    \hline 
         \multicolumn{1}{c}{PM6} \\
         \hline 
         Equilibrium geometries (\AA) \\
         All molecular orbital energies (hartree) \\
         Total electronic and atomization energies (hartree) \\
         \hline
         \multicolumn{1}{c}{$\omega$B97XD/(3-21G, def2SVP, def2TZVP)} \\
         \hline 
         Equilibrium geometries (\AA) \\
         All molecular orbital energies (hartree) \\
         Atomization energies (hartree) \\
         All harmonic frequencies (cm$^{-1}$) \\
         Zero-point vibrational energy (kcal/mol) \\
         Mulliken charges, atomic polar tensor charges ($e$) \\
         Dipole moment (debye) \\
         Polarizability (bohr$^3$) \\
         Radial expectation value (bohr$^2$) \\
         Internal energy at 0~K and 298.15~K (hartree) \\
         Enthalpy at 298.15~K (hartree) \\
         Free energy at 298.15~K (hartree) \\ 
         Total heat capacity (Cal/mol/K) \\
         \hline
         \multicolumn{1}{c}{ZINDO, TD$\omega$B97XD/(3-21G, def2SVP, def2TZVP, def2SVPD)} \\
         \hline 
         Excitation energy of all states (eV, nm) \\
         Oscillator strengths of all excitations (dimensionless) \\
         Transition dipole moment of all excitations (au) \\
    \hline 
    \end{tabular}
    \label{structureddata}
    \end{threeparttable}
\end{table*}

\subsection{Computational Details}\label{sec_compdet}
Initial structures of the 12883 molecules in \bigqm were generated from SMILES by relaxing with the universal force field (UFF)\cite{rappe1992uff} employing tight convergence criteria using OpenBabel\cite{o2011open}.
As a guideline for quantum chemistry big data generation, a previous study proposed connectivity preserving geometry optimizations (ConnGO) to eliminate structural ambiguities due to rearrangements encountered in automated high-throughput calculations\cite{senthil2021troubleshooting}. 
Accordingly, we used a 3-tier ConnGO workflow to generate geometries at the $\omega$B97XD\cite{chai2008long} DFT level using def2SVP and def2TZVP basis sets\cite{weigend2005balanced}. Geometry optimizations at the simpler levels such as PM6 and $\omega$B97XD/3-21G were performed without ConnGO, directly starting from the UFF structures. 
For $\omega$B97XD/def2SVP final geometries, we used HF/STO3G and $\omega$B97XD/3-21G as
intermediate tier-1 and tier-2 levels, respectively. 
Similarly, for $\omega$B97XD/def2TZVP, HF/STO3G and $\omega$B97XD/def2SVP were lower tiers.
In each tier, ConnGO compares the optimized geometry with the covalent bonding connectivities encoded in the initial SMILES and detects molecules undergoing rearrangements. For this purpose, we used the ConnGO thresholds: 0.2 \AA{} for the maximum absolute deviation in covalent bond length 
and a mean percentage absolute deviation of 6\%. 
In DFT calculations, {\tt tight} optimization thresholds and {\tt ultrafine} grids were used for evaluating the exchange-correlation (XC) energy. 
A few molecules required relaxing the optimization thresholds for monotonic convergence towards a minimum. 
All final geometries were confirmed to be local minima through harmonic frequency analysis. 
For molecules that are highly symmetric or with multiple triple bonds, converging to minima was only possible with the {\tt verytight} optimization threshold and {\tt superfine} grids. 
At both $\omega$B97XD/def2SVP and $\omega$B97XD/def2TZVP levels, 3 molecules with the SMILES % Referee 3, minor point 2
{\tt O=c1cconn1}, {\tt N=c1nconn1}, {\tt O=c1nconn1}, failed the ConnGO connectivity tests. Further investigation revealed these molecules to contain an -NNO- substructure in a 6-membered ring facilitating dissociation as previously noted in \RRef{senthil2021troubleshooting}. After removing these  molecules, the size of \bigqm stands at 12880.

We performed vertical excited-state calculations at Zerner's intermediate neglect of differential overlap (ZINDO)\cite{ridley1973intermediate}
and TD$\omega$B97XD levels. 
ZINDO calculations were done on PM6 minimum energy geometries, while TD$\omega$B97XD with 3-21G, def2SVP, and def2TZVP basis sets, at the corresponding ground state equilibrium geometries.
We also performed TD$\omega$B97XD calculations with the diffuse function augmented 
basis set, def2SVPD, on $\omega$B97XD/def2SVP geometries. 
%While the diffuse functions are known to be essential to adequately account for Rydberg excitations, we show using higher-level results for a few test cases that even for the low-lying states, TD$\omega$B97XD/def2SVPD excitation energies and oscillator strengths are more accurate than TD$\omega$B97XD/def2TZVP values.  
%SVP (7s,4p,1d) -> [3s,2p,1d]
%SVPD (8s,4p,2d) -> [4s,2p,2d]
%TZVP (11s,6p,2d,1f) -> [5s,3p,2d,1f]
%TZVPD (12s,6p,3d,1f) -> [6s,3p,3d,1f]
% Referee Points 1 and 3
All electronic structure calculations were performed using the Gaussian suite of programs\cite{frisch2016gaussian}. 
The number of excited  
states accessible to the TDDFT formalism is limited by the number of electrons and
the size of the orbital basis set. With the finite basis set used 
in this study, the spectrum is practically discrete.
To ensure that all the singlet-type
electronic bound states are calculated, we set an upper bound of 10,000 for
the number of states in the TDDFT single point calculations with the keyword {\tt nstates=10000}.
For benchmarking the quality of TDDFT excitation spectra, we also performed similarity transformed equation-of-motion coupled cluster with singles doubles excitation (STEOM-CCSD)\cite{nooijen1997similarity} and the aug-cc-pVTZ
basis set as implemented in Orca\cite{neese2012wiley,neese2018software}. 

\section{Machine Learning Modeling of Full Electronic Spectra}\label{sec:method}
Kernel ridge  regression (KRR) based ML (KRR-ML)
enables accurate predictions
through an exact global optimization of a convex  model\cite{scholkopf2002learning,rupp2012fast,ramakrishnan2017machine}.  
In KRR-ML the target property, $t_q$, of an out-of-sample query, $q$, is estimated as the linear combination of kernel (or radial basis) functions, each centered on a training entry. 
Formally, with a suitable choice of the 
kernel function, KRR approaches the target when the training set is sufficiently large
\begin{equation}
     t_q = \lim_{N \rightarrow \infty} \sum_{i=1}^N c_i k({\bf d}_q-{\bf d}_i).
     \label{eq:mlpred}
\end{equation}
The coefficients, $\{c_i\}$, are obtained by regression 
over the training data.
The kernel function, $k(\cdot)$, captures the similarity in the representations
of the query, $q$, and all $N$ training examples. 
For ground state energetics, 
the Faber-Christensen-Huang-Lilienfeld (FCHL) formalism in combination
with KRR-ML has been shown to perform better than other structure-based
representations\cite{faber2018alchemical,christensen2020fchl}. 
However, for excitation energies and
frontier MO gaps, FCHL's performance drops compared to the 
spectral London-Axilrod-Teller-Muto (SLATM) representation\cite{huang2020quantum}. 
In this study, we compare the performance of FCHL and SLATM 
for modeling the full-electronic spectrum. 
SLATM delivers best accuracies with the Laplacian 
kernel, $k(\textbf{d}_q, \textbf{d}_i) = \exp(-|\textbf{d}_q - \textbf{d}_i|_1/\sigma)$, 
where $\sigma$ defines the length scale of the kernel function and $|\cdot|_1$ denotes $L_1$ norm. 
For the FCHL formalism, we 
found an optimal kernel width of $\sigma=5$ through scanning and a cutoff distance of 20~\AA{}
was used to sufficiently capture the structural features of the longest molecule in the \bigqm dataset, heptane. 

The kernel width, $\sigma$, is traditionally determined through
cross-validation. For multi-property modeling $\sigma$ can be 
estimated using the `single-kernel' approach\cite{ramakrishnan2015many},
where $\sigma$ is estimated as a function of the largest descriptor difference in a sample of the training set, $\sigma = {\rm max}\{ d_{ij} \} /{\rm log}(2)$.
Previous works\cite{ramakrishnan2015many,gupta2021revving,kayastha2020machine} have shown single-kernel modeling to agree with cross-validated results with in the uncertainty arising due to training set shuffles,
especially for large training set sizes. 
KRR with a single-kernel facilitates seamless modeling of multiple molecular properties using standard linear solvers
\begin{equation}
    [{\bf K}+\zeta {\bf I} ] [{\bf c}_1,{\bf c}_2,\ldots] = 
    [{\bf p}_1,{\bf p}_2,\ldots],
    \label{eq:krr}
\end{equation}
where ${\bf p}_j$ is $j$-th property vector and 
${\bf c}_j$ is the corresponding 
regression coefficient vector. We use Cholesky decomposition 
that offers the best scaling of $2N^3/6$ for dense kernel 
matrices of size $N$\cite{van1996matrix}. 
The diagonal elements of the kernel matrix
are shifted by a positive hyperparameter, $\zeta$ to regularize the fit, {\it i.e.,}
prevent over-fitting. 
We note in passing that conventionally the regularization strength is 
denoted by the symbol $\lambda$, which we reserve in this study for wavelength.
Another role of $\zeta$ is to make the
kernel matrix positive definite if there is linear dependency in the feature space arising either due to redundant training entries or due to poor choice of representations. 
Even though we have ensured that our dataset is 
devoid of redundant entries, and the
representations used here accurately map to the 
molecular structure, we cannot {\it a priori}
rule out weak linear dependencies arising from numerical reasons. Hence, we 
condition the kernel matrix by setting $\zeta$ to a small value of $10^{-4}$.
We generated SLATM representation vectors and the FCHL kernel using the QML code\cite{christensen2017qml}, and performed
all other ML calculations using in-house programs written in Fortran. 
All ML errors reported in this study are based on 20 shuffles of the data
to prevent selection bias for small training set sizes. In all learning curves the error bars due to shuffles are vanishingly small for large training set sizes, hence the corresponding envelopes are not shown. Further,
both SLATM and FCHL models were generated using geometries relaxed using UFF in order for the
out-of-sample querying to be rapid.

% In Section~\ref{Section:MLStatespecificitygeom}, we compare 
% the KRR-QML\textcolor{green}{ML} errors for selected ground and excited-state 
% properties with different physical units and 
% magnitudes. To compare the offsets in a learning curve and
% rate of learning across molecular properties, one can use 
% mean percentage errors\cite{kayastha2020machine,gupta2021data}, 
% or by scaling by the maximal value\cite{ramakrishnan2015many}. 
% Both strategies have drawbacks: percentages are not defined when the 
% property values are zero or near-zero, while the maximal value
% is affected by outliers. Hence, in this study, we transform properties
% to their standard scores, $-1\le z \le +1$ as done in data science\cite{igual2017introduction}. 
% The standard score accounts for the sample mean, $\mu_j$ and 
% standard deviation, $s_j$, given as
% \begin{equation}
%     {\bf z}_j = \frac{{\bf p}_j -{\mu_j}}{s_j}.
%     \label{eq:normscore}
% \end{equation}
% Out-of-sample QML\textcolor{green}{ML} errors calculated in units of $s_j$ enable
% meaningful comparisons of error rates across various properties. 

The property (${\bf p}_i$ vector in Eq.~\ref{eq:krr}) modeled in this study corresponds to sum of binned oscillator strength
of electronic transitions from the ground state. 
Conventionally, the band intensity due to the $k-$th excitation 
is the molar absorption coefficient that is proportional to the 
corresponding oscillator strength, 
$f_{0\rightarrow k}$, denoted shortly as $f_k$\cite{turro2017modern}. 
In order to model a full spectrum in a given wavelength range, one
can consider each value of $f_k$ (in atomic units, a.u.) as a separate target quantity. 
However, the number of states is not
uniform in a dataset such as \bigqm. Further, in practice,
one is interested in an integrated oscillator strengths within a small resolution, $\Delta \lambda$. 
Hence, we uniformly divide the spectral range in powers of 2, 
$\Delta \lambda= \lambda_{\rm spectrum}/N_{\rm bin}$, 
where $N_{\rm bin}=1,2,4,\ldots$ is the number of bins. 
For the small organic molecules such as those in \bigqm, we set spectral range to $\lambda_{\rm spectrum}$ to 120 nm capturing most of the excitations. 
For wavelengths $>120$ nm the 
\bigqm dataset provides too few examples, hence, data in this long wavelength
domain is inadequate for ML modeling. 
The target for ML is the sum of $f_k$ in a bin
\begin{equation}\label{eq:p_lambda}
    p_i(\lambda_i) = \sum_{k = 1}^{\rm all\,states} f_{k} (\lambda_k),
\end{equation}
where $i$ is the bin index, and 
$\lambda_i$ is the central wavelength of the bin. 
The oscillator strength of $k$-th excitation from the ground state falls in the 
$i$-th bin if $\lambda_k\in \left( \lambda_i-\Delta \lambda/2, \lambda_i +\Delta \lambda/2  \right]$.
%where $\lambda_i - \Delta \lambda/2 < \lambda_k \le \lambda_i +\Delta \lambda/2$, 
Since we consider $f_k$ in a.u., 
$p_i$ are also in the same units.
% in $1,\ldots,N_{\rm bin}$. 
We explore the performance of ML models for various number of bins.
For the limiting case, $\Delta \lambda= \lambda_{\rm spectrum}$, the target property is the sum of oscillator strengths of all excitations in the selected spectral range, {\it i.e.,} all the intensities are in one bin. The maximum number of bins explored is 128, which results in a spectral resolution of 0.94 nm (=120/128). In this
case, Cholesky decomposition is performed using equation Eq.~\ref{eq:krr}) with 128 columns on the right side, while the number of rows correspond to the training set size.

\paragraph*{Mean absolute error:} 
In this study, our target property is the TD$\omega$B97XD-level binned oscillator strength defined in Eq.~\ref{eq:p_lambda}. Given reference-level TD$\omega$B97XD spectra, the error in the spectra predicted with another model (different theory or ML) 
can be quantified using the standard metric, mean absolute error (MAE):
\begin{equation}
    {\rm MAE} (\Delta\lambda) = 
    \frac{1}{N_{\rm mol}} \sum_{a=1}^{N\rm{_{mol}}}  \sum_{i=1}^{N\rm{_{bin}}} 
  |p^{\rm ref.}_{a,i} - p^{\rm pred.}_{a,i}| ,
  \label{eq:confscore_MAE}
\end{equation}
where $N_{\rm mol}$ is the number of molecules under consideration. 
For a given resolution ({\it i.e.} bin width), $\Delta \lambda$, the error per molecule
is defined by summing the absolute deviations over all bins. For properties such as atomization energy, the desired
target accuracy in MAEs is well-established to be 1 kcal/mol. However, for oscillator strengths of the entire spectra such an
accuracy threshold is not established. Further, relative/percentage errors cannot be defined for oscillator strengths because of the
possibility of vanishing denominators. Similarly, a correlation metric such as the Pearson-$r$ is not
defined for comparing spectra at the limiting case of one bin as it is unreliable for
comparing spectra with fewer bins. Hence, we introduce a new accuracy metric to compare normalized $p_i$ across
two methods and quantify the prediction score on a scale similar to that of percentage error.
\paragraph*{\bf Accuracy metric for normalized spectra:} For the $a$-th molecule, $\tilde{p}_{i,a}$ is the
normalized oscillator strength for the $i$-th bin defined as
$\tilde{p}_{i,a}  = {p}_{i,a} / \sum_i {p}_{i,a}$. For two spectra binned at a common
resolution, $\Delta \lambda$, the accuracy metric for normalized spectra ($\Phi$) is given by:
\begin{equation}
    {\rm \Phi}_a(\Delta\lambda) =
    100\times \left[ 1 - \sum_{i=1}^{N\rm{_{bin}}} 
  |\tilde{p}^{\rm ref.}_{i,a} - \tilde{p}^{\rm pred.}_{i,a}|
  \right].
  \label{eq:confscore}
\end{equation}
When the reference and target property vectors are the same, the accuracy is maximum, $\Phi=100$. For a sample with 
$N_{\rm mol}$ molecule, an overall prediction accuracy ($\overline{{\rm \Phi}}$) can be obtained as an average
\begin{equation}
    \overline{{\rm \Phi}}(\Delta\lambda) = \frac{1}{N_{\rm mol}} \sum_{a=1}^{N\rm{_{mol}}} {\rm \Phi}_a (\Delta\lambda).
  \label{eq:confscore_1}
\end{equation}

\section{Results and Discussions}
\subsection{TDDFT Modeling of Excited States}
A prerequisite for ML modeling is the availability of training data generated with accurate
theoretical levels for the properties of interest. In practice, it is also desirable that the theoretical levels offer a sustainable high-throughput rate for data generation. The more recent
`mountaineering efforts' have reported extended excited states benchmarks of 
highly accurate wavefunction methods for carefully selected 
sets of few hundred molecules\cite{loos2018mountaineering,loos2020mountaineering,loos2020mountaineering2}. Another popular dataset for benchmarking excited state properties was developed by Schreiber \textit{et al.}\cite{schreiber2008benchmarks}. These studies have explored
a wide range of correlated excited state methods including the very accurate 
fourth-order coupled-cluster (CC4) method that approaches the full configuration interaction limit very closely. 
However, even for the lower order methods such as 
third-order coupled-cluster (CC3), excited state modeling becomes
challenging for a large set of molecules.

\begin{figure}[!htbp]
    \centering
    \includegraphics[width=\columnwidth]{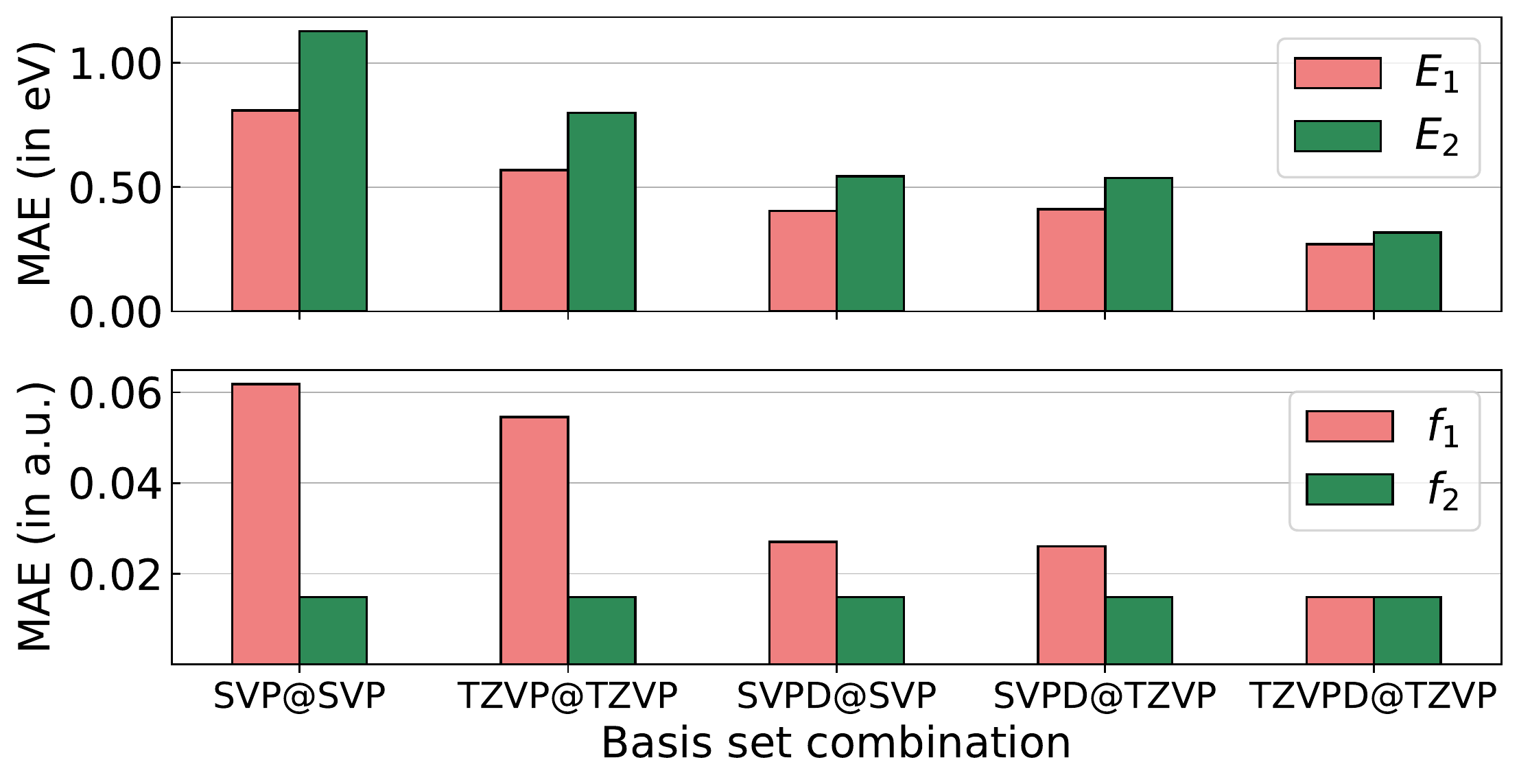}
    \caption{Errors in TD$\omega$B97XD predictions of the
    lowest two excitations, with various basis sets, compared to STEOM-CCSD/aug-cc-pVTZ. Results
    are presented for the smallest 33 molecules in \bigqm with up to three CONF atoms. 
    Mean absolute errors (MAEs) 
    are reported for excitation energies ($E_1$, $E_2$) in the top panel, and
    oscillator strengths ($f_1$, $f_2$) in the bottom panel. 
    The basis sets combination are denoted as: (TDDFT single point)@(DFT structure relaxation).
    %Root mean square deviations are provided above the histograms.
    }
    \label{fig:DFTerror}
\end{figure}
For the low-lying excited states of small molecules, equations-of-motion coupled cluster with singles doubles (EOM-CCSD)\cite{korona2003local} and approximate second-order coupled-cluster (CC2) deliver a mean error of 0.10--0.15 eV compared to higher-level wave function methods\cite{adamo1999toward,laurent2013td,jacquemin2008td,jacquemin2009extensive,loos2018mountaineering,loos2020mountaineering,chrayteh2020mountaineering,loos2020mountaineering2}.
While these methods can be made more economical by using the resolution-of-identity (RI) technique, as in RICC2\cite{send2011assessing} or domain-based local pseudo-natural orbital (DLPNO) variant of EOM-CCSD\cite{berraud2020unveiling}, they
have known limitations when modeling the full electronic spectra of thousands of molecules. 
Formally, the total number of electronic states accounted for by these wave function methods scales as $\mathcal{O}(N_o^2 N_v^2)$, where $N_o$ and $N_v$ are the numbers of occupied and virtual MOs.
Even for a small molecule such as benzene with a triple-zeta basis set, the size of the resulting electronic Hamiltonian is of the order of millions. It is well known that the iterative eigensolvers used for such
large scale problems converge poorly for higher eigenvalues restricting their usage only to the lowest few electronic states\cite{murray1992improved}. 
Hence, as of now, large scale computations of full electronic spectra across a chemical space dataset are amenable only at the time-dependent (TD) DFT-level\cite{casida1998molecular,casida2012progress} that show an
$\mathcal{O}(N_o N_v)$ scaling.

While DFT offers a suitable high-throughput data generation rate, its accuracy for geometries and properties is dependent on the exchange-correlation (XC) functional. 
The chemical space dataset, QM9, was designed using the hybrid generalized gradient approximation (hGGA), 
B3LYP, with 6-31G(2$df$,$p$) basis set because of their use in the G$n$ family of composite wavefunction methods\cite{curtiss2011gn}. 
For thermochemistry energies, B3LYP has an error of 4--5 kcal/mol\cite{curtiss2005assessment}. 
A recent benchmark study\cite{das2021critical} has shown the range-separated hGGAs from the $\omega$B97 family\cite{chai2008long} to have
errors in the 2--3 kcal/mol window; their performance is second only to the G$n$ methods. 
%Compared to B3LYP, the $\omega$B97 methods also deliver refined molecular structures. %SC: Maybe remove this statement as it does not have supporting citations.
While curating the QM9 dataset, the dispersion corrected variant of $\omega$B97, namely $\omega$B97XD, predicted high-veracity geometries less prone to rearrangements in automated high-throughput workflows\cite{senthil2021troubleshooting}.
%SC: Since, while curating the QM9 dataset, the dispersion corrected variant $\omega$B97XD predicted high-veracity geometries less prone to rearrangements in automated high-throughput workflows\cite{senthil2021troubleshooting}, \bigqm entries were minimized using $\omega$B97XD.
Hence, we resort to $\omega$B97XD for geometry optimization and
its time-dependent variant for modeling the complete 
electronic excitation spectra.

The electronic excitations of the molecules in our dataset are
predominantly in the deep-ultraviolet (deep-UV) to X-ray region. Since the popular flavors of TDDFT depend on the adiabatic
approximation where the orbitals are relaxed to first-order as a linear response, they often fail to describe
the electronic wavefunction of high-lying excited states that can substantially differ from that of the ground state\cite{hait2021orbital}. Such effects may be anticipated especially for excitations of long-range charge-transfer character, Rydberg-type\cite{tozer2000determination} or excitations of core electrons\cite{dreuw2003long}. Additionally, electronic states of doubly 
excited character are not accessible to the
linear-response formalism of TDDFT\cite{yanai2004new}. 
%These facts have been established for a few selected transitions of well-known molecules. 
However, as yet, remedies for improving TDDFT for pathological situations have not been tested over chemical space datasets. 
Furthermore, some of the new methods such as the orbital optimized DFT also suffer from algorithmic errors resulting
in variational collapse to a low-lying state\cite{hait2021orbital}.

To probe the effect of basis sets on the TDDFT-level excited state properties,
we selected the smallest 33 molecules with up to 3 heavy atoms
as a benchmark set. 
Accurate modeling of oscillator strengths and high-lying electronic states require 
basis sets augmented with diffuse functions in order to achieve semi-quantitative accuracy. 
Hence, in Fig.~\ref{fig:DFTerror}, we explore $\omega$B97XD's performance for excitation properties computed at def2SVP (SVP), def2TZVP (TZVP), def2SVPD (SVPD), and def2TZVPD (TZVPD). We use
the lowest two excitation energies ($E_1$ and $E_2$) and the corresponding oscillator strengths ($f_1$ and $f_2$)
with the accurate STEOM-CCSD/aug-cc-pVTZ method as the reference. 
Unsurprisingly, def2SVP has the largest errors across all excitation properties followed by 
def2TZVP, def2SVPD, and def2TZVPD. 
Including diffuse functions results in errors that are almost half of those from basis sets devoid of diffuse functions. 
%Further, the basis set used for geometry optimization (def2SVP vs. def2TZVP) seems to have a minor influence on the overall accuracy of the method. 
The errors for all four properties obtained with the def2SVPD basis set
are very similar irrespective of whether the corresponding geometries were determined with
def2SVP or def2TZVP basis sets. 
Even though def2TZVPD offers the best accuracies, we find the computational
cost for determining the full spectra of all molecules in \bigqm to be very high. 
\begin{figure}[!hpb]
    \centering
    \includegraphics[width=\linewidth]{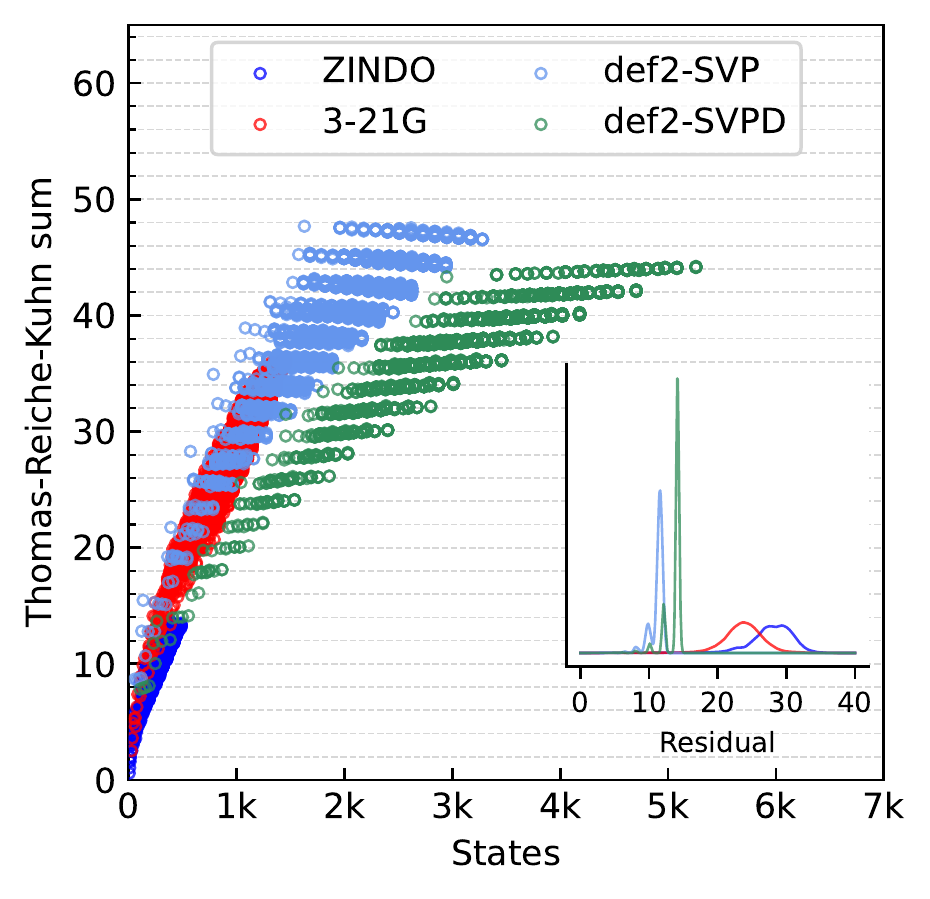}
    \caption{Basis set effect on oscillator strength sums for \bigqm molecules
    at TD$\omega$B97XD. Sum of oscillator strengths of all states is plotted 
    against number of allowed excitations from the ground electronic state with 3-21G,
    def2SVP and def2SVPD basis sets. ZINDO values are also shown for comparison.  
    Horizontal lines mark the Thomas-Reiche-Kuhn limit an exact excited state method must
    coincide with. The inset shows the distribution of deviation of oscillator strength sums
    from total number of electrons.}
    %Referee 3, minor point 3
    \label{fig:TRKlimit}
\end{figure} 
Hence, we resort to the 
def2SVPD basis set that is cost-effective for the excited state calculations. The final target-level
data used for training ML models were obtained at the TD$\omega$B97XD/def2SVPD level using geometries
calculated at the $\omega$B97XD/def2SVP level. 
While TD$\omega$B97XD/def2SVPD level excitation spectra is by no means quantitatively accurate, for high-throughput explorations of medium-sized molecules, it still preserves broad trends that can be learned through structure-property relationships.
%Hence, based on these two benchmarks, we used $\omega$B97XD functional with def2SVP basis set for geometry optimization, and def2SVPD for excited state property estimation for the \bigqm dataset.

We also compare the performance of different methods for predicting 
the Thomas--Reiche--Kuhn (TRK) sum, $\sum_k f_k$.
For an exact excited state method, this sum according to the TRK theorem
must converge to the number of electrons\cite{turro2017modern}. 
In quantum chemistry, unfortunately, this condition is satisfied only 
at the full-CI limit, when all excitations (singles, doubles, triples, and so on) are accounted for at the
basis set limit. ZINDO and the TD$\omega$B97XD methods are not expected to satisfy the
TRK limit. We illustrate this aspect in 
Fig.~\ref{fig:TRKlimit} where the TRK-sum is plotted 
as a function of total number of states accessible. % Referee 3, minor point 4
ZINDO deviates the most from the  TD$\omega$B97XD/def2SVPD target
because the number of excited states available is limited by
two factors. Firstly, 
core electrons are not included in ZINDO. Secondly, semi-empirical models are implicitly based on a 
minimal basis set.

TDDFT modeling with 3-21G improves $\sum_k f_k$
and the total number of states compared to ZINDO. 
With the def2SVP and def2SVPD basis sets, $\sum_k f_k$ quantizes at even numbers with
a separation of about 2. For the large basis set, def2SVPD, the number of  accessible  states increases,
while the TRK-sum drops below the def2SVP values. 
We investigated the reason for this trend using methane and found the
def2- basis sets to show somewhat oscillatory convergence with the basis set size. For methane, 
the def2SVP/def2SVPD values are 8.78/8.12, the larger basis set value 
agreeing better with the aug-cc-pV5Z basis set limit value of 7.82.
%At def2SVPD, the degree of quantization improves with points approaching integer values better.
Residual errors in ZINDO/TDDFT TRK-sums from total number of electrons are 
shown in the inset to Fig.~\ref{fig:TRKlimit}.

\begin{figure}[!htbp]
    \centering
    \includegraphics[width=8.0cm]{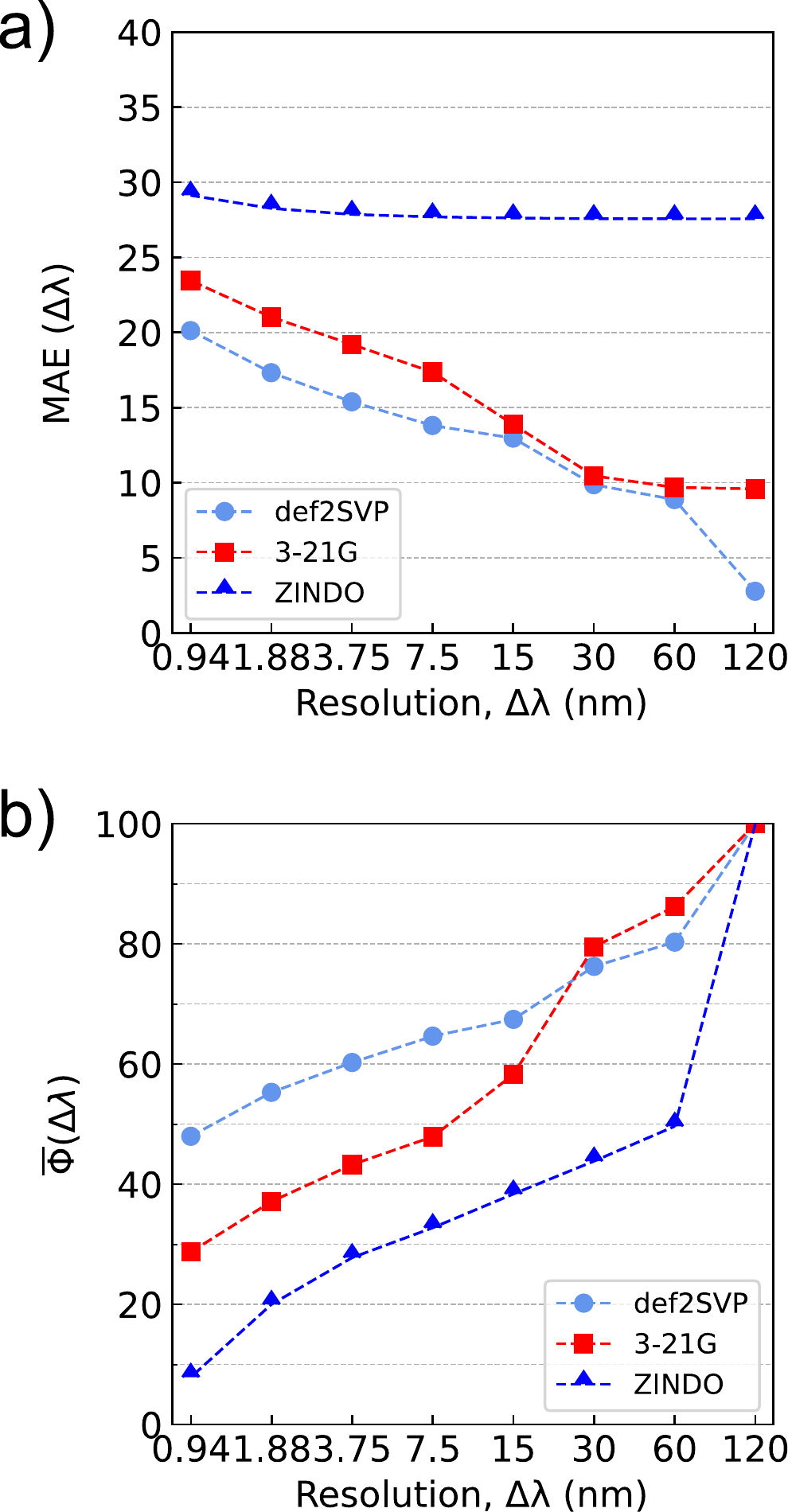}
    \caption{Accuracy metrics for binned oscillator strengths in the $\lambda \le 120$ nm range for
    all molecules in \bigqm: 
    a) Mean absolute error, MAE$(\Delta\lambda)$, in atomic units (a.u.) as defined in Eq.~\ref{eq:confscore_MAE}, 
    b) Mean accuracy metric for normalized spectra, $\overline{{\rm \Phi}}(\Delta\lambda)$, as defined in Eq.~\ref{eq:confscore_1}.
    Results are shown for 
    ZINDO, $\omega$B97XD/3-21G and $\omega$B97XD/def2SVP for approximating $\omega$B97XD/def2SVPD level values.}
    \label{fig:DFTres}
\end{figure}
\subsection{Resolution-vs.-Accuracy Trade-off}\label{sec:DFTcompare}
Typically, uncertainties of hybrid-DFT approximations compared to higher-level wavefunction methods
are used as threshold accuracies for gauging the performance of ML models. 
For the \bigqm dataset, 
along with the conventional MAE metric, we also explore 
a dimensionless accuracy metric for 
normalized spectra, $\Phi$ (see Eq.~\ref{eq:confscore}) and its average. 
Even though the electronic spectra of molecules in \bigqm span a wavelength range until 850 nm, 
$>99\%$ of the spectra lie in the deep UV to X-ray 
range ($10-120$ nm).
Such a trend has been noted before
for small organic molecules\cite{zheng2016electronic}. 
Hence in this study, we fix the spectral range ($\lambda_{\rm spectrum}$) to 0--120 nm 
and bin oscillator strengths at various wavelength resolutions ($\Delta \lambda$) according 
to Eq.~\ref{eq:p_lambda}.  
For a given $\Delta \lambda$, we compare MAE and $\overline{\Phi}$ of predictions from
ZINDO, $\omega$B97XD/3-21G, or $\omega$B97XD/def2SVP levels
with that of the $\omega$B97XD/def2SVPD values (see Fig.~\ref{fig:DFTres}).
For atomization energies and low-lying excitation energies, these
values are 3--4 kcal/mol, and 0.2--0.3 eV\cite{loos2020mountaineering2}, respectively. 
For oscillator strengths, such a threshold has not been established, 
especially for chemical space datasets. 
%For a meaningful comparison of results from different levels, \INBLUE{cancel remaining} we also account for systematic shifts.
% \begin{figure}[!htbp]
%     \centering
%   %\includegraphics[width=8.5cm]{resolution.pdf}
%     \includegraphics[width=\linewidth]{confidence_res.pdf}
%     \caption{Confidence scores (see Eq.~\ref{eq:confscore}) for transition probabilities in the $\lambda \le 120$ nm range for
%     all molecules in \bigqm.  The scores are shown for 
%     ZINDO, $\omega$B97XD/3-21G and $\omega$B97XD/def2SVP levels with for varying wavelength resolution ($\Delta \lambda$) for approximating the
%     $\omega$B97XD/def2TZVP level values. Inset shows the distribution of reference $\omega$B97XD/def2SVPD
%     values.}
%     \label{fig:DFTres}
% \end{figure}

%\textcolor{green}{Hence, the difference between the target and predicted properties were shifted with respect to the mean of the predicted properties.}
% we do not shift the difference by the mean of predicted properties.

%because the total probability is 1 at all levels.  
%The number of bins, $N_{\rm bin}$, is increased in powers of 2 until  sub-nm resolutions are reached. 

In Fig.~\ref{fig:DFTres}a, the MAE of ZINDO shows a smaller variation with $\Delta \lambda$. For the extreme case of 
$\Delta \lambda=120$ nm, where the oscillator strengths of all states are summed in a bin, ZINDO's
MAE saturates to about 27.5 a.u implying a systematic error in ZINDO. For the desired resolution of
0.94 nm, ZINDO's error increases only slightly. The MAEs improve for the spectra calculated with $\omega$B97XD/3-21G.
For the single bin case, the 3-21G results also indicate a systematic error albeit of a smaller magnitude compared to ZINDO. 
The errors are further quenched for the def2SVP basis set, which for a resolution of $\Delta \lambda=0.94$ nm has an
MAE of about 20 a.u. Overall, the MAE-vs-$\Delta \lambda$ dependency becomes stronger in the order: 
ZINDO < $\omega$B97XD/3-21G < $\omega$B97XD/def2SVP. This trend is in agreement with the magnitude of TRK-sum as predicted by these methods, see Fig.~\ref{fig:TRKlimit}. In general, a similar trend is noted also for individual oscillator strengths.

As pointed out in Section-\ref{sec:method}, for the limiting case of $\Delta \lambda=120$ nm,
when all oscillator strengths are summed in one bin, the $\Phi$ is
100 for ZINDO, $\omega$B97XD/3-21G, and $\omega$B97XD/def2SVP methods compared to the target $\omega$B97XD/def2SVPD (see Fig.~\ref{fig:DFTres}b). With increasing resolution, the methods diverge from the target, ZINDO showing the largest deviation from $\omega$B97XD/def2SVPD.
For a desirable resolution of 1\% of $\lambda_{\rm spectrum}$, $\Delta \lambda \thickapprox 1$ nm, 3-21G and def2SVP predictions result in $\Phi$s of $30-50$ compared to the target, while ZINDO
has a worse score $\thickapprox10$. %For a semi-quantitative agreement, with \INBLUE{$\Phi>75$}, ZINDO requires a resolution of 15 nm. 
The reason for poor $\Phi$s of ZINDO 
predictions at small resolution is because core states
are absent in ZINDO, limiting the spectral range to $>19.8$ nm. In contrast, the 
density of the states at the target TD$\omega$B97XD level is high in the
short wavelength domain.
%At the TD$\omega$B97XD-level, when decreasing the quality of the basis set from def2SVPD to 3-21G or def2SVP a \INBLUE{$\Phi$} of $75$ is obtained for a spectral resolution of 8 or 4 nm, respectively, establishing the resolution-vs.-accuracy trade-off.
Applying bin-specific systematic
corrections can improve both the accuracy metrics for all three methods: ZINDO, $\omega$B97XD/3-21G, and $\omega$B97XD/def2SVP.
However, such corrections may not result in uniform improvement throughout the spectral range. For instance, at short
wavelength regions where the TD$\omega$B97XD spectra are sharp, ZINDO lacks these lines. However, 
systematic corrections may result in vanishing MAE for the wrong reason. On the other hand, the effect of such corrections will be less severe for 
the normalized metric, $\overline{\Phi}$. Hence, we do not apply bin-specific systematic corrections in this analyses.
Overall, at the desired resolution of 0.94 nm, among the methods inspected here, the one with larger MAE
has the smaller accuracy metric, $\overline{\Phi}$ and vice versa.

\begin{figure*}[!htbp]
    \centering
    \includegraphics[width=16cm]{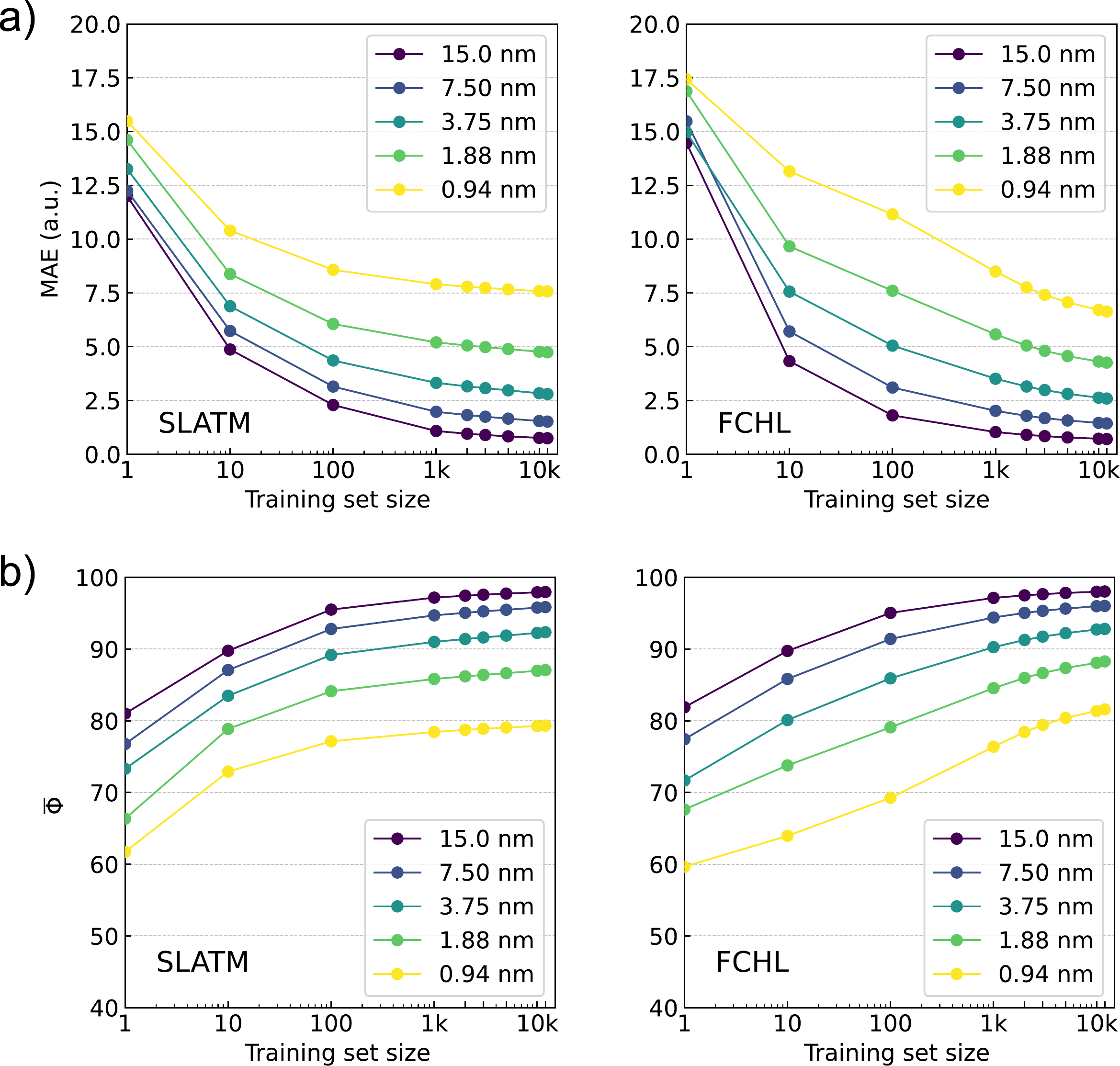}
    \caption{Learning rates based on accuracy metrics for out-of-sample predictions of
    $\omega$B97XD/def2SVPD level 
    binned oscillator strengths (in the $\le 120$ nm region) for the \bigqm dataset:
    Panel a reports MAE in a.u. and Panel b reports $\overline{{\rm \Phi}}$ as functions of 
    training set size for ML models trained using the single-kernel approach with SLATM (left) and FCHL 
    (right) representations generated using UFF-level geometries. 
    %For each curve, the ML target is a vector of length $N_{\rm bin}$, modeled simultaneously with single-kernel. %Horizontal dashed line
    %meets the learning curve for a score of 75\%. 
}
    \label{MLprob}
\end{figure*}
 \subsection{Reconstruction of electronic spectra with ML models}\label{Section:MLStatespecificitygeom}

In Fig.~\ref{MLprob}, we report learning rates based on MAE and $\Phi$ for predicting
binned oscillator strengths (defined in Eq.\ref{eq:p_lambda}) using KRR-FCHL and KRR-SLATM models for various 
training set sizes and spectral resolutions. For poor resolutions, we find small MAEs and large $\Phi$s 
already at the offset of learning curves. 
%At lower resolutions, SLATM achieves convergence faster than FCHL-KRR-ML. 
For large training set sizes and at all resolutions FCHL slightly outperforms SLATM.
While the SLATM model saturates to an MAE of $\thickapprox$7.5 a.u. and $\overline{\Phi}\thickapprox80$ for 0.94 nm resolution 
FCHL model shows improved learning rates, suggesting its scope for modeling full-electronic spectra of 
larger datasets. These findings indicate that it is possible to employ ML modeling for
reconstructing electronic spectra at a high-resolution. Since the ML models were trained on $p_i$ (binned oscillator strengths),
the predicted spectra can be compared with the reference TD$\omega$B97XD spectra similarly binned. The prediction error of the
reconstructed spectrum may be quantified either as a sum of absolute differences, or using the accuracy metric
upon normalizing the binned intensities. The definitions of the error metric do not influence the ML-reconstruction of the
spectra, but they serve merely to quantify the mean prediction accuracy.

The spectra reconstructed with these models do not contain any state-specific information, but rather indicate
the intensity of dipole absorption in a finite wavelength window. 
At the limit of very small $\Delta \lambda$, 
these bins will correspond to individual transitions. 
It is worth noting that for a resolution of 0.94 nm, TD$\omega$B97XD/def2SVP spectra
agree with that of the target-level only with a score of $\thickapprox47$.
The $\Phi$ drops even further for TD$\omega$B97XD/3-21G ($\thickapprox29$) and ZINDO ($\thickapprox9$) levels. 
The learning rates in our evaluatory $\Delta$-ML\cite{ramakrishnan2015big}  calculations using
ZINDO, TD$\omega$B97XD/3-21G, or even TD$\omega$B97XD/def2SVP baseline spectra were inferior 
than modeling directly on the TD$\omega$B97XD/def2SVPD target. Hence, all ML models were trained directly on
the target.

\begin{figure}[!hbtp]
    \centering
    \includegraphics[width=\columnwidth]{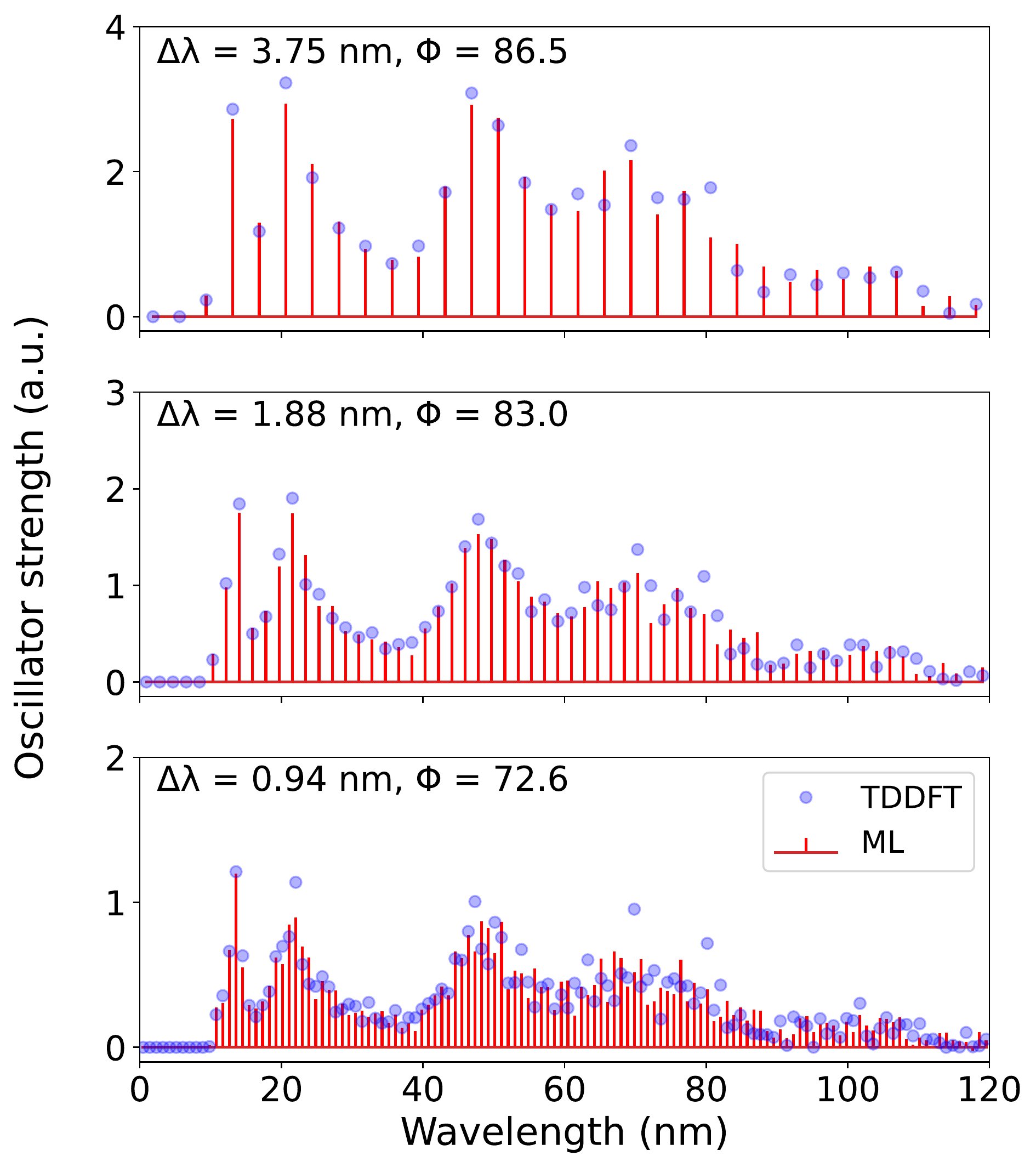}
    \caption{Electronic excitation spectrum of 
    of cyclohexanone,
    reconstructed 
    at 3.75 nm, 1.88 nm, and 0.94 nm resolutions
    using a 1k FCHL-KRR-ML model trained on
    binned oscillator strengths ($p_i$ in Eq.\ref{eq:p_lambda}) at the 
    TD$\omega$B97XD/def2SVPD target-level. Accuracy metric for normalized spectra, $\Phi$,
    compared to TD$\omega$B97XD reference values calculated according to Eq.~\ref{eq:confscore} are also given.
    }
    \label{reconstruction_binning}
\end{figure}

In Fig.~\ref{reconstruction_binning}, we present the entire spectrum of an out-of-sample
molecule, cyclohexanone, reconstructed using FCHL-ML models with 1k training examples at three different wavelength resolutions--3.75 nm, 1.88 nm, and 0.94 nm. 
Since the ML models were trained using geometries at the UFF level, 
these out-of-sample predictions were performed with in a matter of seconds. 
As a part of the supplementary material, we provide 
a sample code for generating the spectrum using a trained FCHL models (see Data Availability). 
For $\Delta \lambda=3.75$ nm, the ML-reconstructed spectrum agrees with the target TD$\omega$B97XD spectrum 
with a $\Phi$ of 86.5. 
This accuracy drops for higher resolutions due to the fine details present
in the target spectrum. Also, with increase in resolution we note a reduction in the spectral heights in order
to conserve the total area under the spectrum. For the desired value of $\Delta \lambda=0.94$ nm, the 
spectrum of cyclohexanone is reconstructed with a score of 72.6 which is slightly lower than the mean score 
reported for out-of-sample predictions in Fig.~\ref{MLprob}. 

% Using a 1k FCHL model based on UFF geometries, we reconstruct the full
% electronic spectrum at the sub-nm resolution $\Delta \lambda=0.94$ nm 
% for four\textcolor{green}{three} random out-of-sample molecules, see FIG.~\ref{reconstruction}.
% In order to compare the individual spectra with the absolute 
% heights from experimental intensities, one can scale the predicted transition probabilities
% by the TRK-sum of each molecule that can be separately predicted by another ML model. 
% For an exact {\it ab initio} method, the TRK-sum should be $N_e$. However, the TDDFT-level
% TRK-sum underestimates this limit due to the absence of double and 
% higher excitations. Hence, to empirically reach the experimental intensities, here we scale the ML predicted transition probabilities by $N_e$. 

\begin{figure}%[!htpb]
    \centering
    \includegraphics[width=\columnwidth]{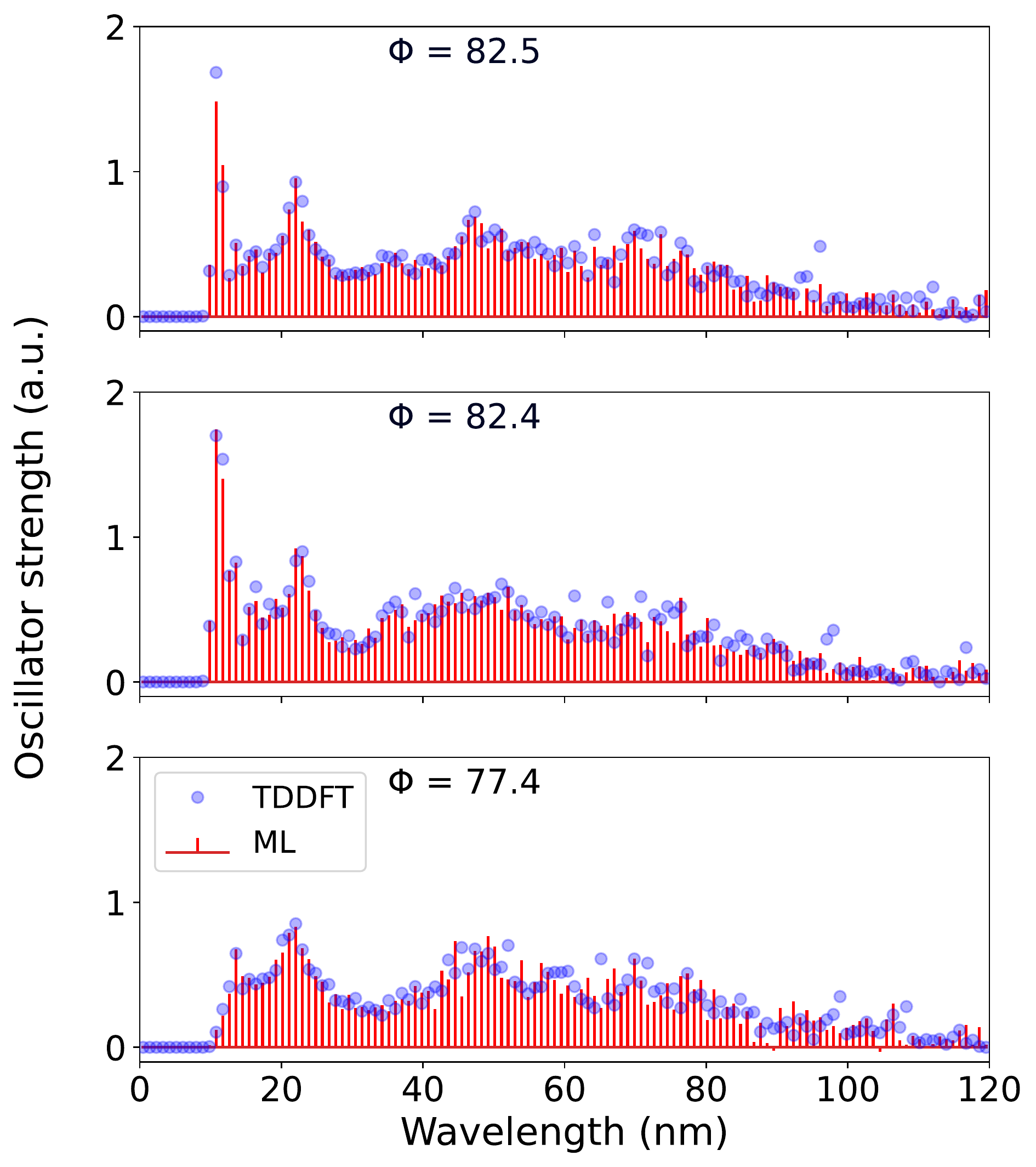}
    \caption{Electronic excitation spectrum of 
    three randomly selected molecules---(3Z)‐5‐fluoro‐4‐methylpenta‐1,3‐diene, 
    1‐fluoropentan‐3‐ol, and 5,5‐dimethyl‐4,5‐dihydro‐1H‐pyrazole --- reconstructed 
    at 0.94 nm resolution using a 1k FCHL-KRR-ML. The model was trained on
    TD$\omega$B97XD/def2SVPD electronic spectra in the $\lambda \le 120$ nm wavelength 
     range. Accuracy metric for normalized spectra, $\Phi$,
    compared to TD$\omega$B97XD reference values calculated according to Eq.~\ref{eq:confscore} are also given.
    % 7382
    % 3936
    % 3843
    % 3719
    }
    \label{reconstruction}
\end{figure}

Further, for the highest resolution explored here, we present the ML reconstructed spectra for three more 
randomly drawn out-of-sample molecules in Fig.~\ref{reconstruction}. For all these molecules, the
prediction is better than for cyclohexanone and are illustrative of the model's mean out-of-sample performance. 
While the reference TD$\omega$B97XD-level binned oscillator strengths are always $>0$, the predicted values are not bound, hence, we notice small negative intensities for 5,5‐dimethyl‐4,5‐dihydro‐1H‐pyrazole. For all four out-of-sample 
molecules considered here, the spectral intensities are low for
$\lambda>100$ nm because of the
corresponding excitations in this region is sparse. 
We believe that ML strategy for spectral reconstruction
reported in this study will hold even at the interesting long-wavelength domain when these models are trained
on adequate examples.

\section{Data-mining in MolDis}
The dataset collected in lieu of this study, justifies an endeavor to make 
it accessible to the wider community. 
While unstructured datasets require an additional step of data extraction, a data-mining platform allows us to rapidly perform multi-property querying and screening.
Our data-mining platform MolDis\cite{moldis} is well-suited to cater to such requirements and hence, we are hosting property-oriented mining platforms for minimum energy ground-state structures of 12880 molecules obtained at the $\omega$B97XD/def2SVP \& $\omega$B97XD/def2TZVP levels at \href{https://moldis.tifrh.res.in/datasets.html}{https://moldis.tifrh.res.in/datasets.html} with both ground-state and excited-state properties.

 \begin{figure*}%[!htpb]
    \centering
    \includegraphics[width=\linewidth]{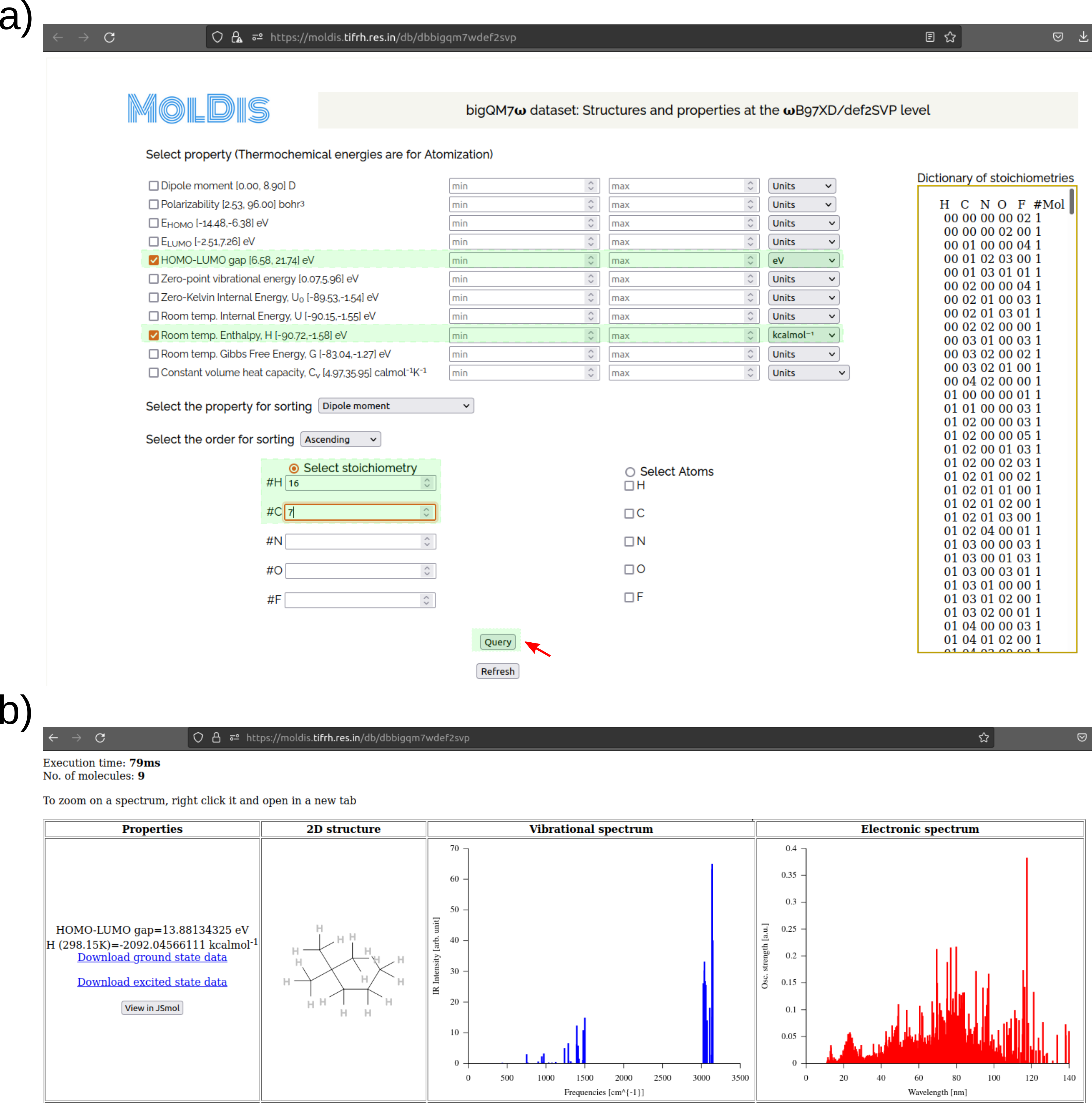}
    \caption{Screenshots of the web-based data-mining platform for querying the
    bigQM7$\omega$ dataset: (a) query page and (b) results page. The example shows how to
    query the HOMO-LUMO gap and room temperature atomization enthalpy of hydrocarbons 
    with the C$_7$H$_{16}$ stoichiometry.
    Separate links are provided at \href{https://moldis.tifrh.res.in/datasets.html}{https://moldis.tifrh.res.in/datasets.html} for accessing minimum energy
    geometries, ground state properties, and vibrational spectra at the $\omega$B97XD/def2SVP and $\omega$B97XD/def2TZVP levels. Electronic spectra calculated at the TD$\omega$B97XD/def2SVPD are also provided.}
    \label{data-mining}
\end{figure*}
In Fig.~\ref{data-mining}, we present a representative property query in the MolDis platform and the corresponding results. 
On accessing the def2SVP tab in the bigQM7$\omega$ {\tt Datasets} page, we arrive at the corresponding query page. 
As noted in Fig.~\ref{data-mining}a, there are 11 ground state properties---dipole moment, polarizability, $E_{\rm HOMO}$, $E_{\rm LUMO}$, $E_{\rm LUMO - HOMO}$, zero-point vibrational energy, zero-Kelvin internal energy ($U_0$), room temperature internal energy ($U$), room temperature enthalpy ($H$), room temperature Gibbs free energy ($G$), and constant volume heat capacity ($C_v$)---with available property ranges reported next to them. 
For a query, users need to enter values with in the property range with appropriate units selected and click on the Query button. 
The search can be further customized upon including multiple properties in the query and display them in ascending or descending order with respect to any property from the corresponding drop-down window.
We have also enabled an option to query based on composition. 
In the bottom half of Fig.~\ref{data-mining}a, users can select either a set of atoms or any valid stoichiometry as listed on the right side of the query page.
Upon making a successful query, users are presented with results (Fig.~\ref{data-mining}b), where the Cartesian coordinates, vibrational and electronic spectra are provided along with the magnitudes of queried properties in desired units. A JSMol applet enables visitors to visualize the structures on their browser upon clicking the ``View in JSMol" button.
Further, upon a fruitful query, both ground-state and excited-state properties for every molecule is presented to the visitor as downloadable files on the results page (Fig.~\ref{data-mining}b).
This platform allows access to \textit{ab initio} properties collected via high-throughput chemical space investigations to the community in a user-friendly fashion, hence, widening the applicability scope of the \bigqm 
dataset.

\section{Conclusions}
In this work we present the new chemical space dataset, \bigqm, containing 12880 molecules with up to 7 atoms of CONF. 
Geometry optimizations of the \bigqm molecules have been performed with the ConnGO workflow
ensuring veracity in the covalent bonding connectivities encoded in their SMILES representation. 
Minimum energy geometries and harmonic vibrational wavenumbers are reported at the accurate, 
range-separated hybrid DFT level $\omega$B97XD using def2SVP and def2TZVP basis sets. This level was selected
because it has been previously shown to result in efficient geometry predictions for chemical space datasets\cite{senthil2021troubleshooting}. We report electronic excited state results at the 
TD$\omega$B97XD level using the def2SVPD basis set containing diffuse functions that are necessary for improved modeling
of oscillator strengths, and high-lying states in general.  
Even for the low-lying excited states of the \bigqm molecules, we found TD$\omega$B97XD/def2SVPD to deliver more
accurate results than the $\omega$B97XD/def2TZVP combination when benchmarked against
STEOM-CCSD/aug-cc-pVTZ reference values. 
For all molecules, full electronic spectra are calculated covering all possible excitations allowed by the TD$\omega$B97XD framework. 
For the small molecules H$_2$O, NH$_3$, and CH$_4$ the resulting number of excited states modeled amounts to 188, 156 and 136, respectively,
while for large molecules such as toluene or $n$-heptane the total number of excited states reported is 
3222 and 5258, respectively. Our preliminary findings have shown that generating the TD$\omega$B97XD results with the even larger basis set
def2TZVPD to require several-fold increase in CPU time. However, when aiming at only a few low-lying states, our results
can be improved when using approximate correlated methods such as DLPNO-STEOM-CCSD(T) or RI-CC2.

For ML modeling of the full electronic spectra, we propose an approach using locally integrated spectral intensities 
at various wavelength resolutions. We illustrate the existence of a resolution-vs.-accuracy dilemma for comparing full electronic spectra from different methods. 
The mapping between the electronic spectra and the global molecular structure-based representations improves only 
when the intensities are binned at a finite resolution.
Semi-quantitative agreement between methods is reached only at the expense of resolution. 
Compared to this, ML models deliver better accuracies at a sub-nm resolution when training on fraction of the dataset. 
For accurate reconstruction of full electronic spectra across chemical space with a resolution of $<1$ nm, we recommend FCHL-KRR-ML. Further, it may be possible to improve the ML model's performance in the long wavelength region using varying resolutions at different spectral regions. 
However, testing this idea requires new datasets comprising adequate data at the desired wavelength domain. 

Our goal is to provide a proof-of-concept for ML modeling of binned electronic spectra and demonstrate accurate spectral
reconstruction.
Unfortunately, the size of the dataset limits the rigor of quantum mechanical methods and basis sets used to estimate the target spectra for ML models. 
While we used range-separated hybrid DFT with moderately large basis sets containing diffuse functions, inherent deficiencies in the method challenge the accuracy of the target.
Further, the small size of the molecules in \bigqm implied excitations modeled are in the far UV region.
However, ML modeling reproduced target spectra at accuracies lower than that arising from deficiencies in the quantum mechanical methods.
This suggests that replacing the target with properties estimated from high-fidelity methods will be adequately captured through ML modeling.

Improvements of ML modeling of excited state requires development of new local descriptors that can map to the chromophores responsible for excitation. 
For this, an automated protocol to characterize electronic excited-states should be developed for high-throughput chemical space design frameworks. 
This allows the opportunity to explore chemically diverse photochemically interesting 
molecules, such as dyes, active in the UV/visible domain and investigate chromophore's/auxochrome's influence on spectra.
Another possibility is to cluster the electronic spectral data according to chromophores \cite{ramakrishnan2015electronic,mazouin2021selected} or by unsupervised learning\cite{cheng2022accurate}.
However, one must ensure that for generating accurate models, each cluster must be adequately represented in the training set. 
In order to facilitate further studies, we provide all data generated for this study in public domains. 
\section{Data Availability}
Structures, ground state properties and electronic spectra of the \bigqm dataset
are available at \href{https://moldis-group.github.io/bigQM7w}{https://moldis-group.github.io/bigQM7w}, see \RRef{kayastha2021Supplementary}.
Input and output files of corresponding calculations are deposited in the NOMAD repository \href{https://dx.doi.org/10.17172/NOMAD/2021.09.30-1}{(https://dx.doi.org/10.17172/NOMAD/2021.09.30-1)}, see \RRef{kayastha2021bigQM7w}. A data-mining platform is available at \href{https://moldis.tifrh.res.in/index.html}{https://moldis.tifrh.res.in/index.html}.

\section{Acknowledgments}
We acknowledge support of the Department of Atomic Energy, Government
of India, under Project Identification No.~RTI~4007. 
All calculations have been performed using the Helios computer cluster, which is an integral part of the MolDis Big Data facility, TIFR Hyderabad \href{http://moldis.tifrh.res.in}{(http://moldis.tifrh.res.in)}.

\section{Author Contributions}
PK and RR conceptualised the project and methodology. 
SC and RR were involved in writing, reviewing, and editing the manuscript.
SC and RR maintain the project content in GitHub and MolDis.
All authors were involved in data generation/curation, analysis and visualization.
Software development, resource/funding acquisition were done by RR.
RR supervised PK and SC.

\section*{Conflicts of interest}
``There are no conflicts to declare''.

\bibliography{ref}
\end{document}